\documentclass[twocolumn,aps,prb,final,superscriptaddress]{revtex4}

\usepackage{graphicx}

\newcommand{\Tl}{Tl$_2$Mo$_6$Se$_6$}
\newcommand{\In}{In$_2$Mo$_6$Se$_6$}

\newcommand{\M}{$M_2$Mo$_6$Se$_6$}
\renewcommand{\deg}{^\circ}
\newcommand{\degC}{$^\circ$C}

\begin{document}

\title{Lattice dynamics of the cluster chain compounds \M}


\author{Liam Gannon}
\affiliation{Diamond Light Source, Harwell Campus, Didcot OX11 0DE, United Kingdom}
\affiliation{Clarendon Laboratory, University of Oxford Physics Department, Parks Road, Oxford, OX1 3PU, United Kingdom}

\author{Lilia Boeri}
\affiliation{Dipartimento di Fisica, Sapienza Universit\`a di Roma and Istituto dei Sistemi Complessi, ISC-CNR,  00185 Rome, Italy}

\author{Christopher A. Howard}
\affiliation{Department of Physics \& Astronomy, University College London, London WC1E 6BT, United Kingdom}

\author{Patrick Gougeon}
\affiliation{Laboratoire de Chimie du Solide et Inorganique Mol\'eculaire, UMR CNRS 6226 - INSA, Universit\'e de Rennes 1, Avenue du G\'en\'eral Leclerc, 35042 Rennes-Cedex, France}

\author{Philippe Gall}
\affiliation{Laboratoire de Chimie du Solide et Inorganique Mol\'eculaire, UMR CNRS 6226 - INSA, Universit\'e de Rennes 1, Avenue du G\'en\'eral Leclerc, 35042 Rennes-Cedex, France}

\author{Michel Potel}
\affiliation{Laboratoire de Chimie du Solide et Inorganique Mol\'eculaire, UMR CNRS 6226 - INSA, Universit\'e de Rennes 1, Avenue du G\'en\'eral Leclerc, 35042 Rennes-Cedex, France}

\author{Diala Salloum}
\affiliation{Permanent address: Faculty of Science III, Lebanese University, PO Box 826, Kobbeh-Tripoli, Lebanon}
\affiliation{Laboratoire de Chimie du Solide et Inorganique Mol\'eculaire, UMR CNRS 6226 - INSA, Universit\'e de Rennes 1, Avenue du G\'en\'eral Leclerc, 35042 Rennes-Cedex, France}

\author{Alexander P. Petrovi\'{c}}
\altaffiliation{Present address: Division of Physics and Applied Physics, School of Physical and Mathematical Sciences, Nanyang Technological University, 21 Nanyang Link, Singapore 637371}
\affiliation{Department of Quantum Matter Physics, University of Geneva, 24 Quai Ernest-Ansermet, 1211 Geneva, Switzerland}

\author{Moritz Hoesch}
\affiliation{Diamond Light Source, Harwell Campus, Didcot OX11 0DE, United Kingdom}
\affiliation{DESY Photon Science, Deutsches Elektronen-Synchrotron, Notekestrasse 85, 22607 Hamburg, Germany,
moritz.hoesch@desy.de}

\begin{abstract}
The lattice dynamics of members of the \M\ family of materials with guest ions $M=$~K, Rb, Cs, In, and Tl has been studied using inelastic x-ray scattering and Raman spectroscopy at room temperature, as well as by {\em ab-initio} calculations. We find a good match between calculations and experiment, both for structure factors (Eigenvectors) and for the calculated phonon frequencies. The observed lattice dynamics for \Tl\  show no signs of anharmonicity or absence of avoided dispersion crossings, thus ruling out previously hypothesised rattling phonon modes. The reduced mode energies for \In\ are identified as only partially responsible for the lower superconducting transition temperature $T_c$ in this material when compared to \Tl.  
\end{abstract}

\maketitle

\section{Introduction}

Materials with uniaxial anisotropy of the crystal lattice are of interest for the research on one-dimensional electronic systems that could form if the charge carriers were confined to motion along individual chains of atoms, molecules or clusters. The electronic structure of real materials will include a finite inter-chain coupling, but it is still justified to speak of a quasi-one-dimensional (q1D) state if the hopping along chains is much easier than between chains. Many materials have been identified as having a suitable chain arrangement in their crystal structure.~\cite{monceau12} These materials thus allow the study of the intriguing and well-established theoretical predictions, in particular if the band filling allows for a metallic state.~\cite{giamarchibook} One-dimensional confinement of the electrons occurs in materials that still feature three-dimensional long-range order and, apart from strong mechanical anisotropies,~\cite{ribeiro2002,gannon2015} have regular vibrational properties.

\begin{figure}
	\centerline{\includegraphics[width=0.46\textwidth]{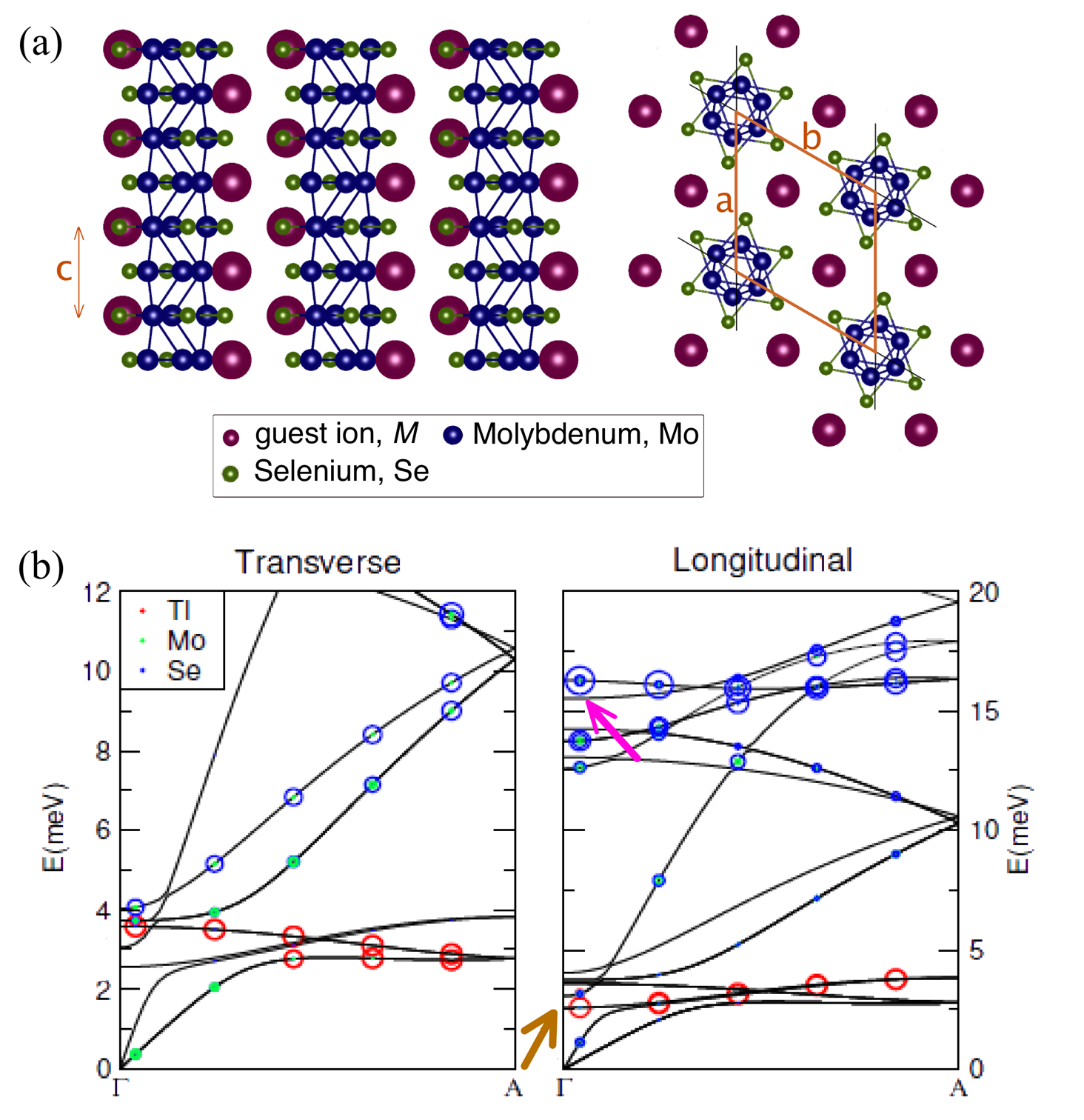}}
	\caption{(a) Crystal structure of \M, seen along $a$ (left) and along the unique axis $c$ (right). (b) Calculated phonon dispersions for \Tl. Coloured circles mark the contributions from Tl, Mo and Se for the transverse (left) and longitudinal (right) phonons. The longitudinal guest-ion and  chain modes are marked by a brown and a purple arrow, respectively. Note the different energy scales for the two panels.}
\label{Fig1}
\end{figure}

The family of compounds \M\ derives from Mo$_3$Se$_3$ cluster units that are stacked along the $c$-axis of a hexagonal lattice with two cluster units per unit cell as shown in Fig.~\ref{Fig1}.~\cite{potel1980,chevrel1985} $M$ guest ions that can be alkali metals (Na, K, Rb, Cs) or group III metals (In or Tl) occupy highly symmetric positions between the chains. Owing to the winding arrangement of Mo$_3$ triangles that are rotated by $60\deg$ from cluster to cluster, the chains are sometimes referred to as nano-tubes. A single electronic band leads to metallic conductivity. This band, derived from Mo 4$d$ states that are de-localised over the clusters, has a chiral wave function as identified by density-functional-theory calculations.~\cite{petrovic10} The inter-chain coupling  increases along the group of alkali metals from Rb to Na and is significantly higher for \In\ and \Tl.~\cite{petrovic2016} This chiral electron band was later described as an electron band of cubic Dirac Fermion type by considering all symmetries of the P6$_3$/m space group,~\cite{gibson2015,liu2017} which was subsequently discussed in terms of topological superconductivity.~\cite{huang2018}

Strongly anisotropic superconductivity (SC) has been observed in all family members, which is remarkable given the highly q1D electronic system that renders the SC unconventional. In \Tl\ this has been consistently observed ever since the first samples were synthesised.~\cite{armici1980} The highest observed transition temperatures  are up to $T_c= 5.8$~K in \Tl  (onset at 6.8 K)~\cite{armici1980} and $T_c= 2.85$~K in \In.~\cite{petrovic10} This SC is likely to be phonon mediated and remarkable thermoelectric properties relate to electron phonon coupling (EPC) as well.~\cite{mori1984,tseng1992,verebelyi1997}  Due to the high temperatures required for synthesis and single crystal growth the samples sometimes have significant amounts of vacancies in the guest ion positions, while the (Mo$_3$Se$_3$) cluster chains have extremely low defect densities.  The stoichiometry can deviate from the ideal \M\ as much as $M_{1.7}$Mo$_6$Se$_6$ for lighter $M$-ions.~\cite{tarascon84,petrovic2016} Alkali metal family members show a broad minimum of resistivity with a minimum at temperatures that increase with increasing disorder.~\cite{petrovic2016} Careful experiments have found SC also in the high-resistance low-temperature state and this was related to localisation effects.~\cite{ansermet2016,petrovic2016}  A dimensional crossover at temperatures $T_X \simeq 4-5$~K in \Tl\ further supports the q1D nature of the electronic structure.~\cite{mitra2017,bergk2011} Nevertheless, an expected Peierls-like instability of the system~\cite{andersen_private,liu2017} has not been found yet and the absence of a structural phase transition is remarkable given the highly nested Fermi surface.

The family members of \M\ share structural similarities with the unconventional superconductors K$_2$Cr$_3$As$_3$,~\cite{bao2015,kong2015} Rb$_2$Cr$_3$As$_3$~\cite{tang2015} ($T_c = 6.1$~K and 4.8~K, respectively), and KCr$_3$As$_3$~\cite{mu2017} ($T_c=5$~K), as well as with TlFe$_3$Te$_3$~\cite{wang2018} (ferromagnetic, no SC). These compounds  have a different valence count and the compounds with 233 stoichiometry have a different arrangement of guest ions and may lack inversion symmetry, which is present in the 133 compounds \M, KCr$_3$As$_3$ and TlFe$_3$Te$_3$. The lattice dynamics of K$_2$Cr$_3$As$_3$ has been studied theoretically~\cite{subedi2015} and a few phonon modes of particularly strong coupling have been found. A Raman scattering study of K$_2$Cr$_3$As$_3$ found evidence for such EPC in the observation of Fano line shapes of the spectra but no temperature-dependent softening or peak splitting was observed.~\cite{zhang2015} 

In the heavy guest ion members of \M\  with $M$= Tl, In and Rb  the EPC of the low-lying vibrational modes have been previously analysed with a combination of transport and specific heat measurements,~\cite{petrovic10} which compares well to inelastic neutron scattering results on polycrystalline samples.~\cite{brusetti90} Rather sharp Einstein modes were found at low energy that related to the guest ion vibrations as identified by the scaling of their energy with the guest ion mass. For these low energy guest ion modes an anomalous difference of EPC between In and Tl compounds was identified and invoked to explain the difference in transition temperatures in \Tl\ and \In\  ($T_c =4.2$~K and 2.85~K, respectively~[\onlinecite{petrovic10}]). The SC coupling is, however, principally provided by higher energy modes of the metallic chains (Mo$_3$Se$_3$)$_\infty$, whose EPC has not yet been determined in fine detail.~\cite{petrovic10} 

In this paper, we report a new study of the momentum resolved lattice dynamics, measured by inelastic x-ray scattering (IXS) as well as by Raman scattering. Two types of modes are of principal interest: the low energy guest ion modes and the higher energy Mo$_3$Se$_3$ chain modes. They are seen (for the example of 
\Tl) as bands in Fig.~\ref{Fig1}(b) between $2-5$~meV for the guest ion modes (strongly depending on guest ion mass) and $12-18$~meV for the chain modes (nearly independent of the guest ion). A survey across a large selection of family members allows us to identify guest ion and pure chain modes and verify with very high precision the scaling found previously.~\cite{brusetti90} We find a good agreement between calculations and experimental results, while deviations from the general trends are seen in \In. The lowering of the higher energy chain mode frequency in \In\ contributes to explaining the large difference in superconducting transition temperature $T_c$. 

This paper is structured as follows: the next section will describe experimental and numerical methods. Section~\ref{sectResults} will summarise the findings that are discussed in Section~\ref{sectDiscussion}; the conclusions drawn from these are summarised in Section~\ref{sectConclusions}.

\section{Methods}
\label{sectMethods}

Needle-shaped samples with hexagonal cross section and thicknesses between 50 and 100~$\mu$m and up to 3~mm length were grown in sealed molybdenum or silica crucibles at temperatures between $1100\deg$C (\In) and $1700\deg$C (\Tl) by similar methods as described previously.~\cite{potel1980} Details of the sample growth are given in Appendix~\ref{appendixSampleGrowth}.
 
IXS spectra were measured at beamline ID28 at the European Synchrotron Radiation Facility (ESRF).~\cite{sette2006} This instrument features a high resolution monochromator using a near-backscattering Si(9 9 9) Bragg reflection, a beam focussing system for a spot size of $60 \times 200 \mu$m (vertical $\times$ horizontal) and a multiplexing analyser and detector system, again utilising the Si(9 9 9) Bragg reflection. In this configuration the combined energy resolution is 3.2~meV.  Part of the data were taken at higher energy resolution of 1.6~meV in the Si(11 11 11) configuration of the spectrometer. The resolution function of the spectrometer was measured in the same data acquisition campaign by scattering from amorphous acrylic glass. The samples were aligned in a Laue geometry with the $c$-axis along the momentum transfer $\vec{Q}$. In this geometry the orientation of $a^*$ and $b^*$ axes (undetermined) is irrelevant and a phonon dispersion along $c^*$ is measured for longitudinal phonons with Eigenmodes along $c$. The data are analysed by a numerical peak fitting that includes the central peak and pairs of phonon-related peaks at positive and negative energy transfer, weighted in intensity to the expected ratio at the given temperature ($T=300$~K). Raw spectra of IXS from \Tl\ are shown in Fig.~\ref{Fig2}(a) as well as in Appendix~\ref{appendixIXSrawdata}.

  IXS dispersions $E(Q)$ were extracted from the spectra at each momentum point $Q$ by numerical fitting of the raw spectra with a model consisting of shifted (along energy loss) and scaled (along intensity) copies of the measured resolution functions. For each phonon mode that is present in the spectra (one or two in the spectra of our study) a pair of resolutions functions with energy losses $\pm E$ models the energy-loss and energy-gain features of the spectra. The relative intensities within these pairs take into account the detailed balance according to bosonic mode occupation at the known temperature (room temperature) and $E$. An additional scaled but unshifted resolution function is added to model quasi-elastic scattering. A fully summed spectrum of this kind is allowed to shift along the energy loss axis to compensate for small drifts of the zero point of the axis that can easily occur in IXS. The fit model thus has $2n+2$ parameters, where $n$ is the number of modes in the spectrum. The reported error bars are numerical uncertainties that derive from the fitting algorithm. An example of selected raw data is presented in Fig.~\ref{Fig2}(a) along with examples of resolution functions shifted in this manner for a single mode.

Raman scattering spectra were measured in a Renishaw inVia micro-Raman spectrometer equipped with a 632 nm laser. The laser beam was focussed to an area of $\approx 3 \ \mu$m and the power on the sample was kept below $2$~mW. The scattered light was analysed using a polarisation filter before admission into the spectrometer. A box filter was used to suppress the intensity of elastically scattered light, which leads to a lowest analysable phonon energy of 12~meV. The data from 20 subsequent detector exposures of 20~s dwell time were summed after removal of spurious intensities from the individual exposures. 

{\em Ab-initio} calculations of the lattice dynamics were performed within Density Functional Perturbation Theory (DFPT),~\cite{DFPT}
as implemented in \textsc{Quantum espresso}.~\cite{QE-2009}
We employed non-relativistic norm-conserving Martins-Troullier pseudopotentials,~\cite{PSP:Martin-Troulliers}
with non-linear core-corrections, from the standard \textsc{Quantum espresso} distribution,
with a plane-wave cutoff of 80 Ryd; the exchange and correlation energy
was treated in the the Generalized Gradient Approximation (GGA),
in the Perdew-Burke-Ernzerhof  parametrization.~\cite{PBE}

For reciprocal space integration, we employed a 
$2 \times 2 \times 8$ regular grid in $\mathbf{k}$-space, and a Methfessel-Paxton
smearing parameter of 0.02 Ryd; phonons were computed on a $2 \times 2 \times 4$
regular grid in reciprocal space, and dispersions on a finer mesh were then obtained by Fourier interpolation of the force constants.
All calculations were carried out at the experimental lattice constants,
reported in Ref.~[\onlinecite{petrovic10}].
In order to have a more accurate estimate of the frequency of the soft phonon modes at the
$\Gamma$ point, the phonon frequencies were recomputed with all-electron
LAPW calculations,~\cite{LAPW_OKA_1975,Wien2k} 
with a much denser ($8 \times 8 \times 24$) $\mathbf{k}$-grid and improved tetrahedron method,~\cite{tetra}
using the frozen-phonon approach. All calculated modes at $\Gamma$ are listed in table S1 in Appendix~\ref{appendixModeTable}.
  The calculated phonon dispersions along the $c^*$ direction for \Tl\ are shown in Fig.~\ref{Fig1}(b), where the left part highlights transverse modes
(i.e. with Eigenmodes orthogonal to $c$) and the right side highlights  longitudinal modes. Modes with dominant contribution from one of the three ions of the compound are marked with colours.

\section{Results}
\label{sectResults}

\begin{figure*}
	\centerline{\includegraphics[width=0.41\textwidth]{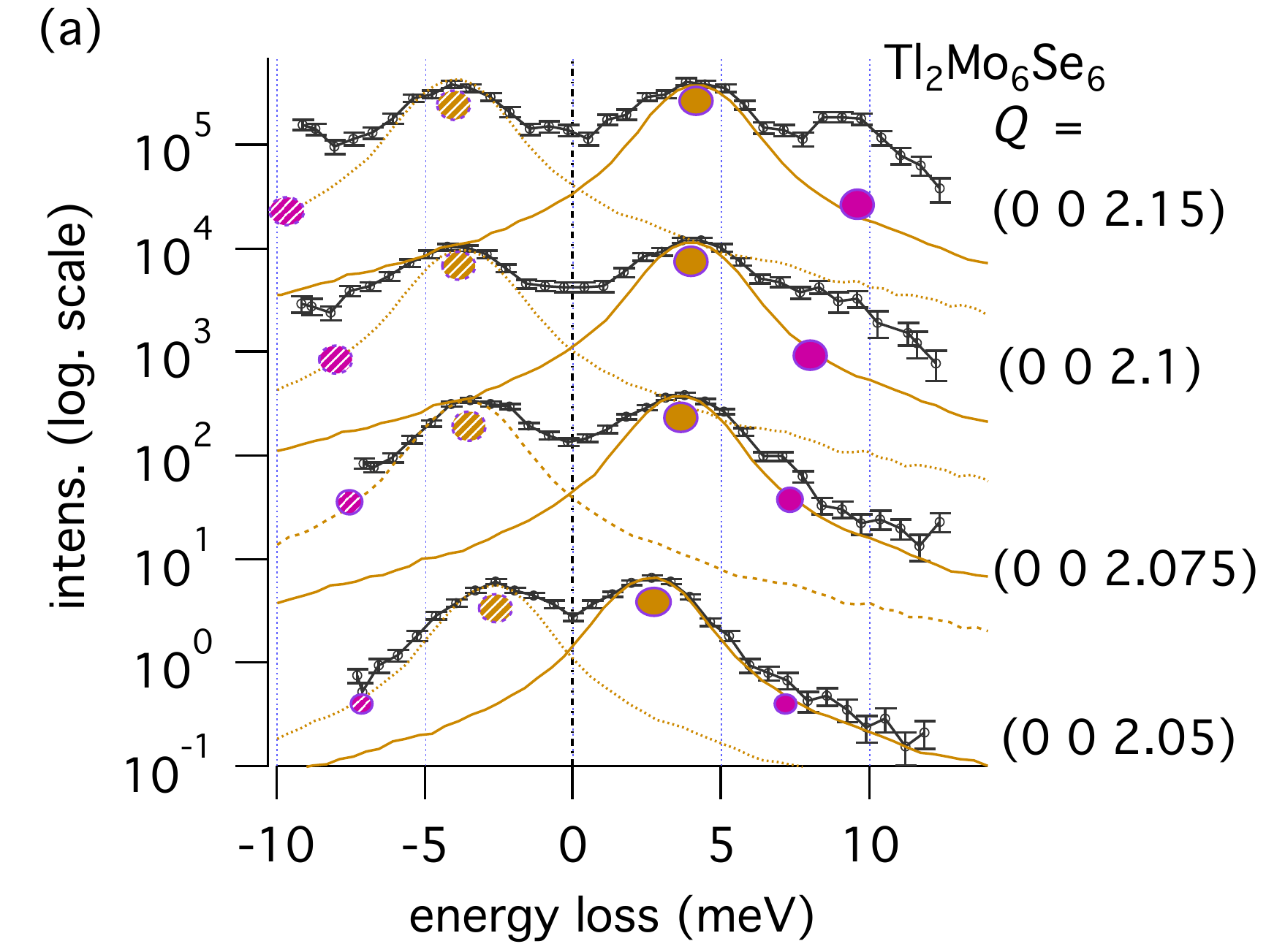}\hspace{3mm}\includegraphics[width=0.57\textwidth]{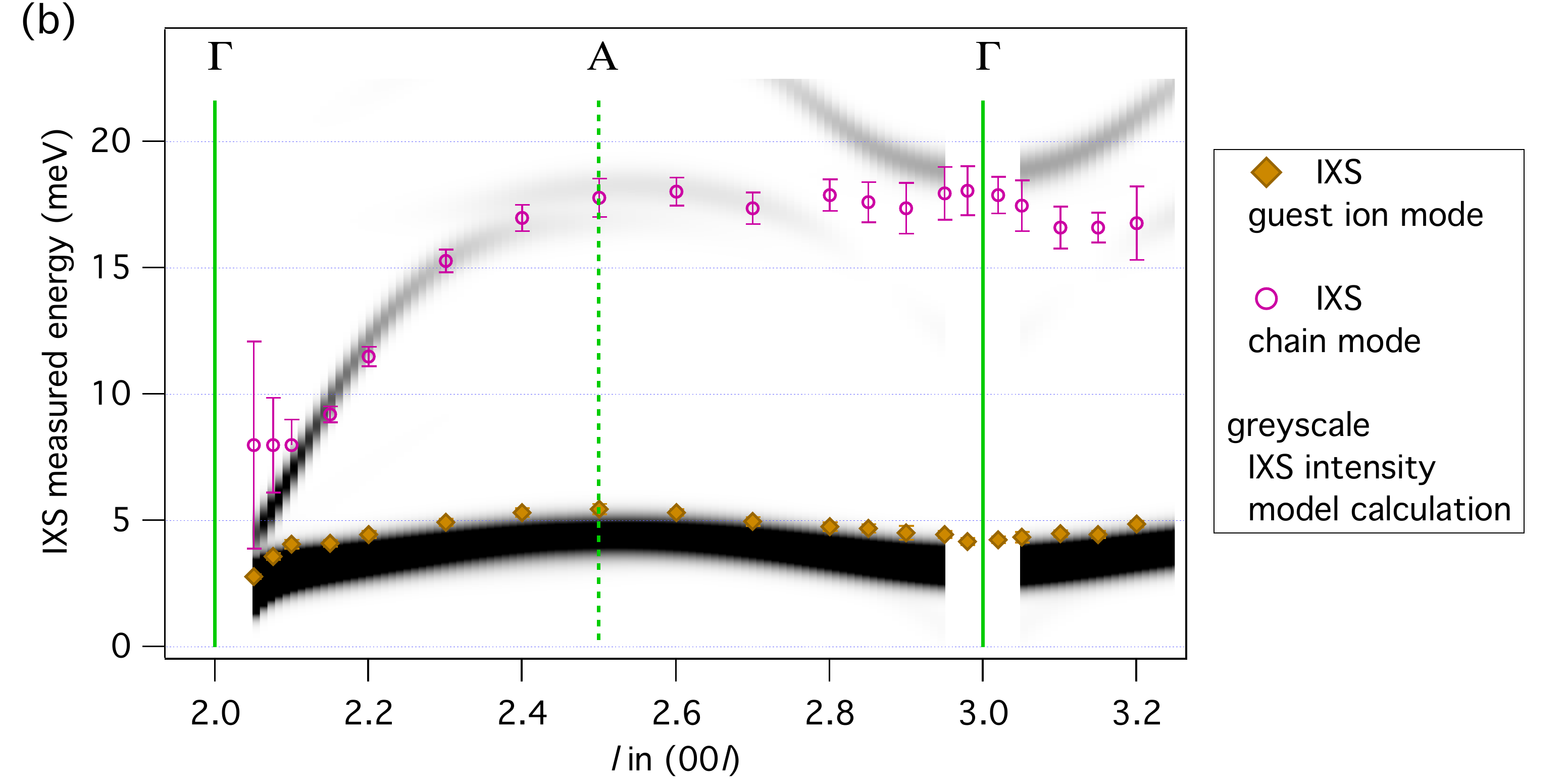}}
	\caption{  (a) Selected IXS spectra of \Tl\ close to the $\Gamma$-point (002). Two observed modes are indicated by brown and purple colour in solid symbols for positive energy loss and dashed symbols for negative energy loss. The resolution function, shifted to match the lowest, strongest phonon mode is shown in two copies in each spectrum. The spectra are offset vertically for graphical clarity. (b) Extracted  phonon dispersions along the $(00l)$-direction for longitudinal  modes of  \Tl. The data are overlaid on a grey-scale representation of the expected IXS active modes in our scattering geometry.  Error bars represent the statistical uncertainty of peak positions from the numerical fits applied to the raw spectra.}
\label{Fig2}
\end{figure*}

We first report the results from the momentum-resolved IXS. Fig.~\ref{Fig2} shows the measured phonon dispersions for two modes up to $E=25$~meV for \Tl. The momentum axis spans from $\vec{Q}=(0,0,2.0)$ to $(0,0,3.3)$. Only longitudinal modes contribute to the signal in this geometry.  (002) is a strong Bragg reflection and corresponds to a $\Gamma$-point, as does the Bragg-forbidden (003). The Brillouin zone boundary point A is reached at $(0,0,2.5)$. The data are overlaid on a grey-scale representation of predicted scattering intensities from the lattice dynamics calculation. The intensities were calculated by methods used in the ab2tds software.~\cite{ab2tds}

At all probed momenta, the strongest IXS intensity from \Tl\ arises from a mode with energies between $2.5 - 7$~meV. This is a guest ion mode with similar but distinct Eigenvectors on either side of the Brillouin zone boundary A, thus giving rise to different IXS intensities near (002) and near (003). From the strong Bragg spot (002) the  steep dispersion of the acoustic mode is observed, which bends into the flat guest ion mode. Near (003) and along the full length of $\ell = 2.5  - 3$, the guest ion mode is a nearly flat pure optical mode. The spectral features of this mode are sharp, resolution limited peaks, thus no sign of strong anharmonicity is observed. 

A second peak at higher energy is found in the data between $E=7 - 17$~meV. The comparison to the calculated IXS intensities shows that this is a well-defined single phonon mode only in small regions close to the $\Gamma$-points. At intermediate momenta, namely around A, a superposition of intensities from related but separate phonon modes contributes to the intensity. Near (002) the steep dispersion bends into an optical mode of $E=8$~meV (with large error bars), which shows that an avoided crossing is present to the acoustic branch of guest ion character.  This is readily seen in the raw data of Fig.~\ref{Fig2}(a) as a double peak feature or excess scattering intensity beyond the strong lowest energy peaks. Near (003) the longitudinal mode can be identified and is indicated by an orange marker in Fig~\ref{Fig1}(b). This is a nearly pure chain mode as shown below.

Fig.~\ref{Fig3}(a) expands the results of Fig.~\ref{Fig2}(b) with data from a large number of different guest ions. As expected, the guest ion mode energies are different near (003), while the chain mode dispersion is identical for all measured family members within the error bars ($\pm2 $~meV), thus proving its character without much guest ion involvement. For the lightest measured guest ion K, no guest ion mode could be detected. For the \In\ sample, a few spurious extra peaks were detected in some of the spectra; these are marked by symbols without error bars. 

\begin{figure*}
	\centerline{\includegraphics[width=.96\textwidth]{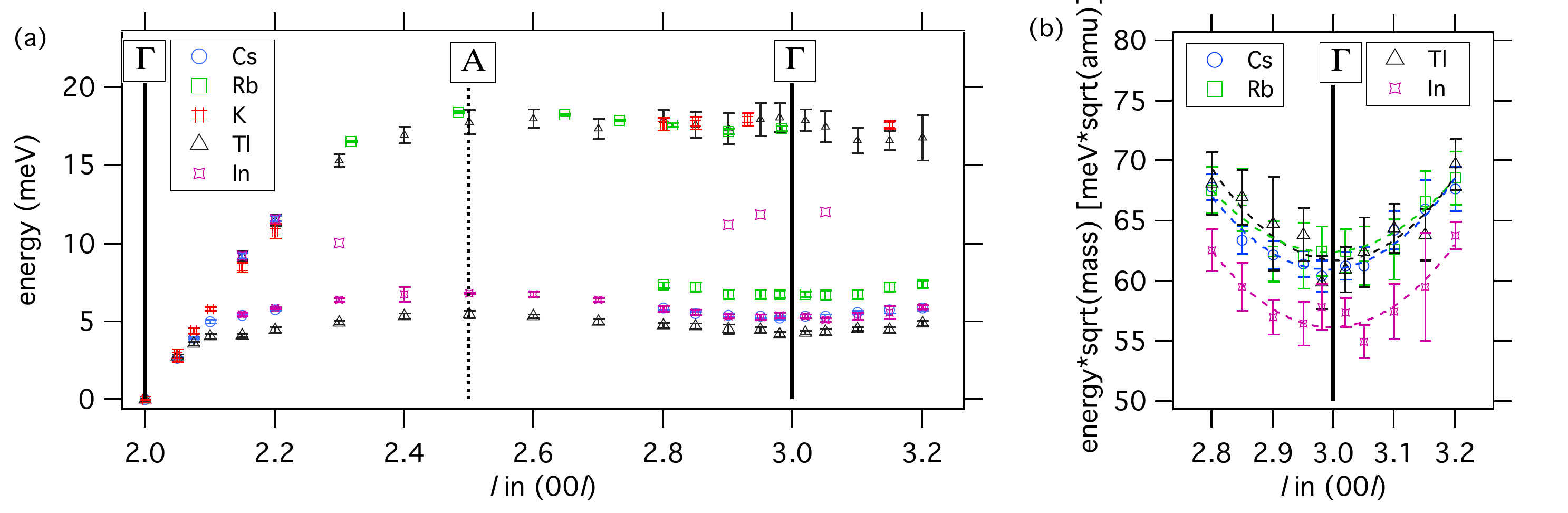}}
	\caption{{(a) Longitudinal phonon dispersions for several \M. (b)} A comparison of the dispersion of the guest-ion modes about the $\Gamma$-point, normalised against the guest-ion mass.}
\label{Fig3}
\end{figure*}

The momentum region very close to the $\Gamma$-point (003) is shown in Fig.~\ref{Fig3}(b). It is rare and fortunate to be able to measure IXS spectra so close to $\Gamma$ as strong Bragg reflection intensity can dominate the spectra and preclude analysis. At (003) Bragg scattering is suppressed, which highlights the good quality of the crystals for all family members. Well-defined peaks with a clear dispersion are connected by parabolic fits to the data for each material. The data have been rescaled by division with the square root of the respective guest ion mass.~\cite{webelements} In agreement with earlier neutron scattering data~\cite{brusetti90} the mass-corrected phonon energies fall precisely on top of each other, both in terms of $\Gamma$-point energy as well as in terms of dispersion. This identifies the mode as a pure guest ion mode without any contribution of the chains. One exception is found in \In. Here the dispersion agrees with the other compounds, but the $\Gamma$-point energy is lower by 8\%.

\begin{figure*}
	\centerline{\includegraphics[width=\textwidth]{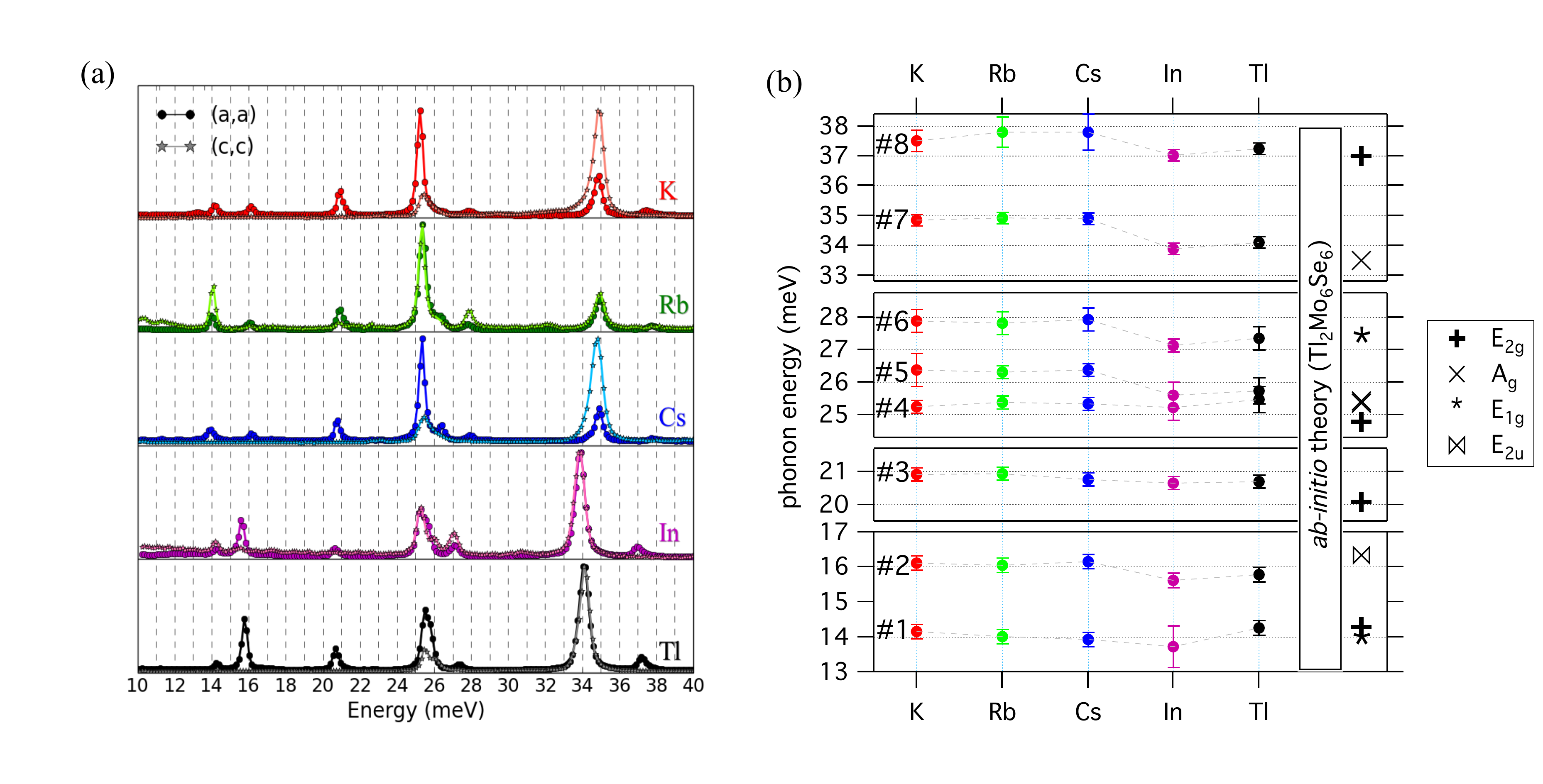}}
	\caption{(a) Raman spectroscopy of the low-energy phonon modes in the series \M. Spectra with the light polarisation perpendicular and along the crystal $c$-axis are shown as circles and stars, respectively. (b) Peak positions of all modes observed in (a) as a function of guest ion.}
\label{Fig4}
\end{figure*}

Raman scattering spectra are shown in Fig.~\ref{Fig4}(a) for compounds with five different guest ions. The lowest detectable Raman excitation was $E=14$~meV in the range of chain modes, below this the necessary box filter removes the intensity. Data were measured with the polarisation filter adjusted to either parallel to the crystal $c$-axis (aligned by eye to the needle-like shaped sample) or orthogonal to it. The polarisations of incoming and outgoing photons were thus $(c,c)$ or $(a,a)$, respectively. While selection rules are not consistently giving rise to observed or suppressed peaks in all samples, the majority of peaks are visible only in $(a,a)$ polarisation. In this geometry, several rotational modes of Mo$_3$Se$_3$ triangles are expected to contribute to the Raman signal. The longitudinal modes that were observed in IXS are not Raman active. 

We can now identify the Raman active modes with two methods. Purely experimentally we can verify the scaling with the guest ion mass, namely modes with identical energy for all family members are identified as pure chain modes. For \Tl\ the comparison to the {\em ab-initio} calculation further confirms this assignment. Fig.~\ref{Fig4}(b) shows the analysed peak positions for eight distinct peaks. In the range of $10-30$ meV, we identify features \#2, as well as \#5 and \#6 as mostly pure chain modes by their identical energies for K, Rb and Cs. These modes are of E$_{2u}$, A$_u$ and E$_{1g}$ symmetry with Mo/Se in-plane (\#2 and \#5) and out-of-plane (\#6). Modes \#1 and \#3 show a trend of decreasing phonon energy with increasing guest ion mass across the series of Alkali metals (K, Rb, Cs) and are identified as E$_{2g}$ modes with Mo/Se out-of-plane and in-plane, respectively. Peak \#1 may furthermore have a contribution from a Raman-active E$_{1g}$ mode that is too close in energy to the E$_{2g}$ to be resolved in the spectra.

\section{Discussion}
\label{sectDiscussion}

The data of a large number of members of the \M\ family give a consistent picture of the lattice dynamics of both the guest ion and chain modes as well as collective modes. Specifically for \Tl, the observed data compare to {\em ab-initio} calculations with a good agreement. This agreement together with the resolution-limited peak widths and avoided crossings near $\Gamma$, shows that none of the guest ion modes can be considered as rattling phonon modes. The lattice dynamics is truly collective at all momenta including the guest ion modes. 

The re-scaling to atomic masses of the guest ion, first reported in Ref.~[\onlinecite{brusetti90}] is re-confirmed with much higher resolution thanks to being able to probe the longitudinal mode very close to $\Gamma = (003)$ in isolation. This confirms identical scaled phonon energies and thus a high purity of this  guest ion mode. One exception is present in \In, which has a lower energy guest ion vibration. Previous studies did not have the resolution to find this difference in scaled mode energy, but found a significantly weaker EPC for this mode in \In\ when compared to \Tl.~\cite{petrovic10} Surprisingly, our new data show a softening of this mode in \In\ when compared to the other family members, while a reduced EPC naively suggests that this mode should be hardened.

The chain modes, observed in Raman spectroscopy also show a deviation of \In\ from the trends of the other family members. For several Raman active modes the phonon energies for \In\ are significantly reduced when compared to the other family members. These chain modes provide most of the coupling responsible for SC~\cite{petrovic10} and their reduced energy leads to an effective reduction of the prefactor in the McMillan formula for superconductivity,~\cite{mcmillan1968} partially explaining the reduced superconducting $T_c$. The actual difference in $T_c$ between \Tl and \In, which can be as high as a factor two, depending  on samples, is however much larger than the observed softening, which is of the order of $5-10\%$; the remaining difference must come from a different coupling to the low-energy guest ion modes, and/or to reduced matrix elements.

Also, \Tl\ show slightly lower chain mode energies than expected from the trend across the whole series of Raman data. The measured energies are well reproduced by the calculation within 1~meV, with some modes higher and some lower than observed. Note that in all compounds the Raman peaks are sharp and show no visible sign of Fano line-shapes at room temperature, thus these data do not provide direct evidence for a particularly strong EPC with no exceptions from material to material.

\section{Conclusions}
\label{sectConclusions}

The data of IXS and Raman scattering, combined with the {\em ab-initio} calculations  for \Tl\  give the consistent result that all members of the \M\ series, feature regular lattice dynamics as can be expected for compounds of this kind. The guest ion motion is harmonic and Einstein-like pure guest ion branches merge into collective modes with the expected anti-crossing   (observed in \Tl). Therefore, a rattling-mode description of the guest ions is not supported by the data. The data also show no indication of a soft mode that could be related to a Peierls-like phase transition, neither from the comparison of samples with varying inter-chain coupling nor by momentum dependent softening signatures in any given sample.

\In\ falls out of the series because it exhibits both a reduced rescaled mode energy of the guest ion optical mode, when compared to the other family members, as well as reduced energies of several chain modes. \Tl\ also has some reduced energies of these modes. This indicates that the difference between $p$-electron charge transfer for In and Tl, compared to $s$-shell transfer for alkali metals influences not only the electronic inter-chain coupling~\cite{petrovic2016} but also the bonding in the chains leading to softened chain modes. The stronger softening in \In\ when compared to \Tl\ partially explains the significantly lower superconducting $T_c$.

\acknowledgments
We wish to thank O.~K. Andersen and N. Ogita for fruitful discussions and A. Bosak and R. Burkovsky for technical assistance. Beamtime at beamline ID28 of the ESRF contributed to the results shown here. The Raman apparatus was provided by the Central Laser Facility of STFC. One of us (MH) would like to thank Hiroshima University and the Bodeleian Libraries for their kind hospitality during the time of writing this manuscript. 

\appendix

\makeatletter  
\renewcommand{\fnum@figure}{FIG. S\thefigure}
\renewcommand{\fnum@table}{TABLE S\thetable}
\makeatother     

\section{Sample synthesis and single crystal growth}
\label{appendixSampleGrowth}
Starting materials used for the solid-state syntheses and crystal growths were MoSe$_2$, InSe or TlSe, MCl (M = K, Rb, and Cs) and Mo, all in powder form. In order to avoid any contamination by oxygen and moisture, the starting reagents were kept and handled in a purified argon-filled glovebox. Before use, Mo powder was reduced under H$_2$ flowing gas at 1000\degC\ during ten hours in order to eliminate any trace of oxygen. The molybdenum diselenide was prepared by the reaction of selenium with H$_2$ reduced Mo in a ratio 2:1 in an evacuated (ca. 10$^{-2}$~Pa Ar residual pressure) and flame-baked silica tube, heated at about 700\degC\ during two days. InSe and TlSe were synthesized from the elements in evacuated sealed silica tube at 800\degC\ and 400\degC, respectively, for 1 day. Powder samples of the $M_2$Mo$_6$Se$_6$ ($M$ = K, Rb, and Cs) compounds were prepared in two steps. The first one was the synthesis of In$_2$Mo$_6$Se$_6$ from a stoechiometric mixture of InSe, MoSe$_2$ and molybdenum heated at 1000\degC\ in evacuated sealed silica tube for 36 hrs. The second step was an ion exchange reaction of In$_2$Mo$_6$Se$_6$ with MCl (M = K, Rb, and Cs) at 800\degC. For the latter one, powders of In$_2$Mo$_6$Se$_6$ and MCl in a ratio of about 1:2.5 were mixed and then cold pressed. The pellet was subsequently sealed under vacuum in a long silica tube. The end of the tube containing the pellet was placed in a furnace with about 5 cm of the other end out from the furnace, at about the room temperature. The furnace was heated at 800\degC\ for 48 h. After reaction, white crystals of InCl are observed at the cool end of the tube. The resulting product was found to be single-phase on the basis of its powder X-ray diffraction diagram made on a D8 Bruker Advance diffractometer equipped with a LynxEye detector (Cu K$_{\alpha 1}$ radiation). Single crystals of the alkali metal members $M_2$Mo$_6$Se$_6$ ($M$ = K, Rb, and Cs) as well as those of Tl$_2$Mo$_6$Se$_6$ were obtained by heating cold pressed powder samples (ca 5g) into a molybdenum crucible sealed under a low argon using an arc-welding system. The different charges were heated at the rate of 300 \degC/h up to 1500 \degC\ (1700\degC\ for Tl$_2$Mo$_6$Se$_6$), temperature which was held for 3 hours, then cooled at 100 \degC/h down to 1000\degC\ and finally furnace cooled. Crystals thus obtained have the shape of needles with hexagonal cross section. Single crystals of In$_2$Mo$_6$Se$_6$ were obtained by heating powder samples in evacuated sealed silica tube at 1150\degC\ for 72 h.

\section{Calculated Mode table}
\label{appendixModeTable}

The phonon frequencies at $\Gamma$ were computed with all-electron LAPW calculations, with a  dense ($8 \times 8 \times 24$) $\mathbf{k}$-grid and improved tetrahedron method using the frozen-phonon approach. The resulting phonon mode frequencies, symmetries and information about infrared or Raman active modes are given in Table~\ref{mode_table}.

\begin{table*}[h]
\caption{Calculated phonon modes of \Tl\ at $\Gamma$, their symmetry and Infrared or Raman activity (Act.): I = IR active; R = Raman active. Markers (Mark) as in Fig.~4 of the main paper.}
\label{mode_table}
\begin{center}
\begin{tabular}{|ccccc||ccccc|}
\hline
\hspace{1mm}Omega\hspace{1mm} &  \hspace{1mm} Omega\hspace{1mm} &   \hspace{1mm}  Sym. \hspace{1mm}  &   \hspace{1mm}Act.\hspace{1mm} &\hspace{1mm}Mark\hspace{2mm} &\hspace{1mm}Omega \hspace{1mm}& \hspace{1mm}  Omega\hspace{1mm} &  \hspace{1mm}   Sym. \hspace{1mm} &   \hspace{1mm}Act.\hspace{1mm}&\hspace{1mm}Mark\hspace{1mm} \\  
(cm$^{-1}$) &   (meV)  &  & & & (cm$^{-1}$) &   (meV)  &  &  & \\
\hline
0.0   & 0.0 &     E1$_u$* &  I &  & 175.4 &  21.7456 & A$_u$ &  I     & \\ 
0.0 &   0.0 &     A$_u$ &   I  & & 179.6 &  22.2663 & B$_g$ &        & \\
20.7 &  2.56633 & B$_g$ &   & &   189.7 &  23.5185 & B$_u$ &         & \\ 
24.4 &  3.02504 & A$_u$ &   I  &  & 196.2 &  24.3243 & E$_{1u}$ &  I     & \\ 
28.8 &  3.57054 & E$_{2g}$ &  R &   &  196.2 &  24.3243 & E$_{1u}$* &  I    & \\  
28.8 &  3.57054 & E$_{2g}$* &  R  &    & 199.9 &  24.783 & E$_{2g}$ &  R  & \bf{+}   \\
29.9 &  3.70692 & E$_{1u}$ &  I   &   &  199.9 &  24.783 & E$_{2g}$* &  R  & \bf{+}\\
29.9 &  3.70692 & E$_{1u}$* &  I  &    & 204.6 &  25.3657 & A$_g$ &  R   &\bf{$\times$} \\ 
32.4 &  4.01686 & A$_g$ &   R   &   &  205.2 &  25.4401 & A$_g$ &  R    & \bf{$\times$}\\  
105.5 &  13.0796 & B$_u$ &     &   &  219.3 &  27.1882 & E$_{1g}$ &  R   &\bf{*} \\  
109.1 &  13.5259 & B$_g$ &      &  &    219.3 &  27.1882 & E$_{1g}$* &  R  & \bf{*}\\
110.9 &  13.7491 & E$_{1g}$ &  R  & \bf{*}  & 261.9 &  32.4696 & E$_{2u}$ &     & \\     
110.9 &  13.7491 & E$_{1g}$* &  R  & \bf{*} &    261.9 &  32.4696 & E$_{2u}$* &      & \\   
115.1 &  14.2698 & E$_{2g}$ &  R   & \bf{+}  &    265.2 &  32.8788 & B$_u$ &       & \\    
115.1 &  14.2698 & E2g* &  R  & \bf{+} & 266.2 &  33.0027 & E$_{1u}$ &  I    & \\
125.2 &  15.5219 & E$_{1u}$ &  I    &   &   266.2 &  33.0027 & E$_{1u}$* &  I     & \\ 
125.2 &  15.5219 & E$_{1u}$* &  I &  &  270.2 &  33.4986 & A$_g$ &  R  & \bf{$\times$}     \\  
131.8 &  16.3402 & E$_{2u}$ &    & $\bowtie$  & 275.8 &  34.1929 & B$_u$ &    & \\
131.8 &  16.3402 & E$_{2u}$* &  & $\bowtie$ & 298.5 &  37.0072 & E$_{2g}$ &  R   & \bf{+} \\ 
162.0 &  20.0843 & E$_{2g}$ &  R  & \bf{+} &   298.5 &  37.0072 & E$_{2g}$* &  R   &\bf{+} \\
162.0 &  20.0843 & E$_{2g}$* &  R  &\bf{+} &           &               &       & & \\
\hline
\end{tabular}
\end{center}
\end{table*}

\section{Raw IXS spectra}
\label{appendixIXSrawdata}

\begin{figure*}
	\centerline{\includegraphics[width=0.88\textwidth]{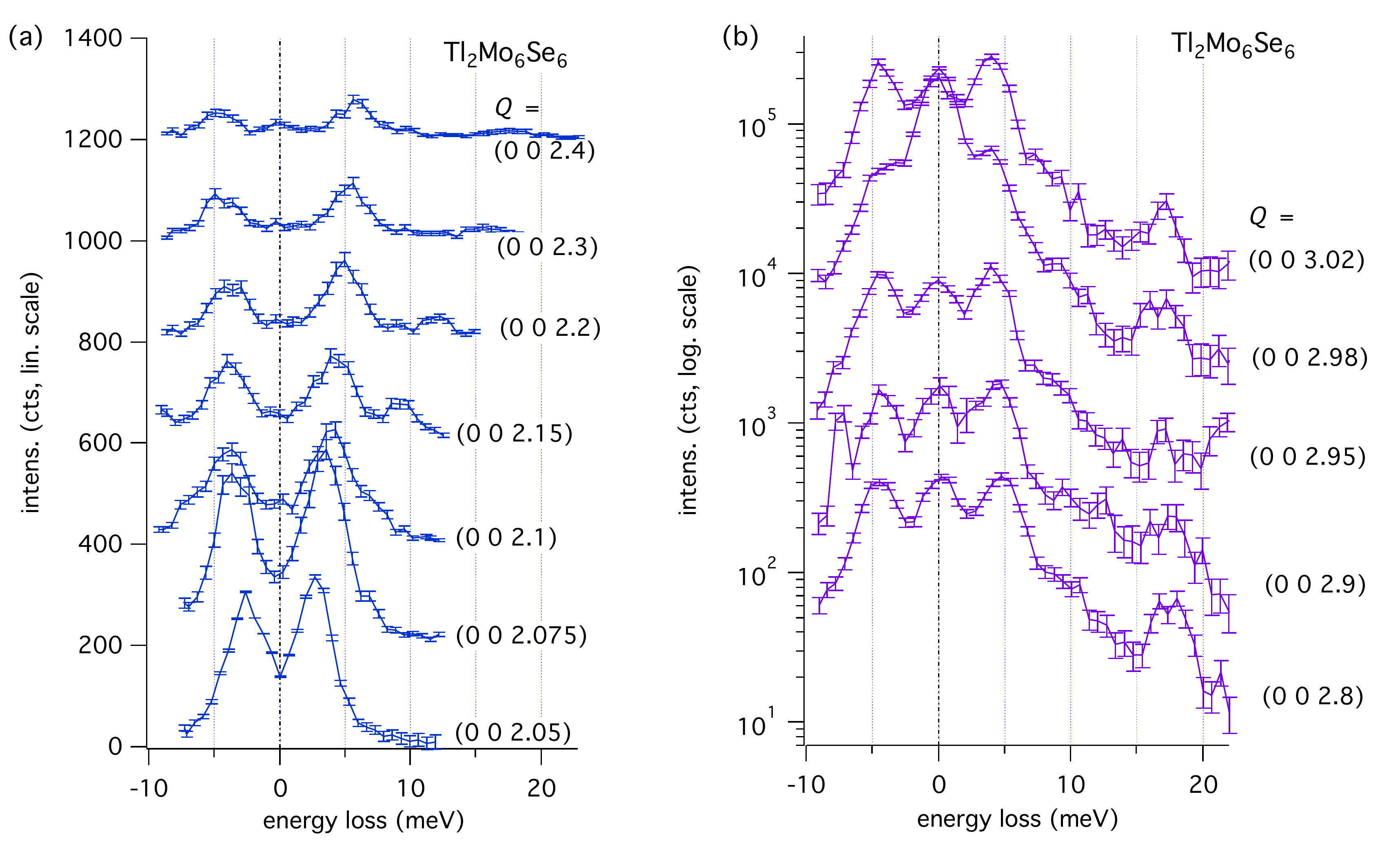}}
	\caption{ Phonon spectra from \Tl\ measured at the indicated momentum position $\vec{Q}$. The spectra are offset vertically for clarity. }
\label{Supp_Fig1}
\end{figure*}

\begin{figure*}
	\centerline{\includegraphics[width=0.88\textwidth]{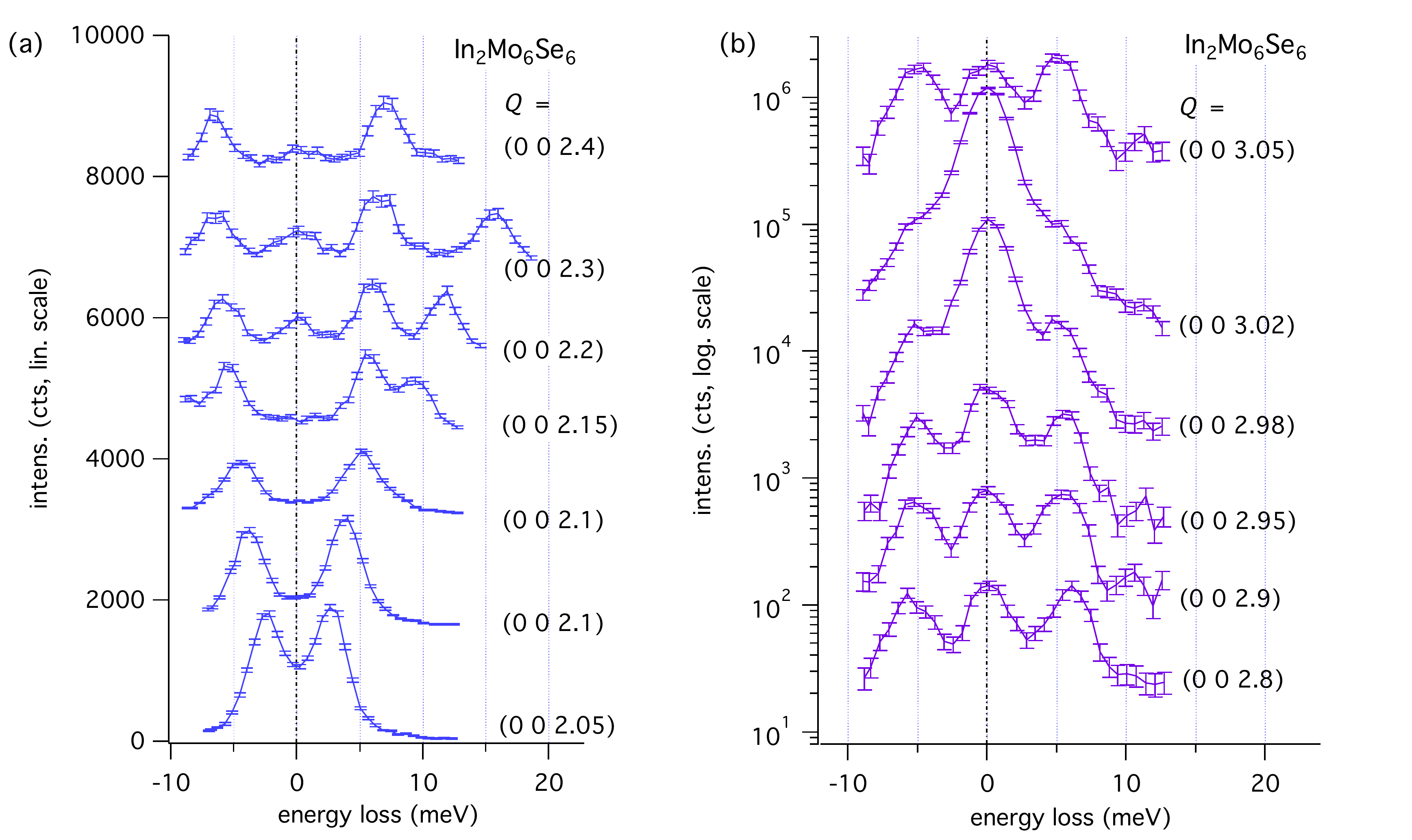}}
	\caption{ Phonon spectra from \In\ measured at the indicated momentum position $\vec{Q}$. The spectra are offset vertically for clarity. }
\label{Supp_Fig2}
\end{figure*}

Raw spectra of inelastic x-ray scattering (IXS) from \Tl\ and \In\ are shown in Figs.~S\ref{Supp_Fig1} and S\ref{Supp_Fig2}, respectively. The momentum transfer $\vec{Q}$ is indicated for each spectrum. Spectra in Figs.~S\ref{Supp_Fig1}(a) and~S\ref{Supp_Fig2}(a)were acquired in 30 seconds per point, spectra in Figs.~S\ref{Supp_Fig1}(b) and~S\ref{Supp_Fig2}(b) were acquired for 120 seconds per point. Some spectra shown in Fig.~S\ref{Supp_Fig1}(a) in linear scaling are identical to the ones shown in Fig.~2(a) of the main paper in logarithmic scaling.

\bibliography{Lattice_dynamics}

\end{document}